\documentclass[sigconf]{acmart}

\usepackage{url}
\usepackage{float}
\usepackage{hyperref}

\AtBeginDocument{%
  }

\copyrightyear{2025}
\acmYear{2025}
\setcopyright{cc}
\setcctype{by}
\acmConference[CHI '25]{CHI Conference on Human Factors in Computing Systems}{April 26-May 1, 2025}{Yokohama, Japan}
\acmBooktitle{CHI Conference on Human Factors in Computing Systems (CHI '25), April 26-May 1, 2025, Yokohama, Japan}\acmDOI{10.1145/3706598.3714019}
\acmISBN{979-8-4007-1394-1/25/04}

\begin{document}

\title[Health, Privacy, and Security in IoT Monitoring for Older Adults]{\lq\lq Watch My Health, Not My Data\rq\rq~:~Understanding Perceptions, Barriers, Emotional Impact, \& Coping Strategies Pertaining to IoT Privacy and Security in Health Monitoring for Older Adults}

\author{Suleiman Saka}
\affiliation{%
 \institution{University of Denver}
 \city{Denver}
 \state{Colorado}
 \country{USA}}
 \email{suleiman.saka@du.edu}

\author{Sanchari Das}
\affiliation{%
  \institution{George Mason University}
  \city{Fairfax}
  \state{Virginia}
  \country{USA}}
\email{sdas35@gmu.edu}

\renewcommand{\shortauthors}{Saka and Das}

\begin{abstract}   
 

The proliferation of \lq\lq Internet of Things (IoT)\rq\rq~ provides older adults with critical support for \lq\lq health monitoring\rq\rq~ and independent living, yet significant concerns about security and privacy persist. In this paper, we report on these issues through a two-phase user study, including a survey ($N=22$) and semi-structured interviews ($n=9$) with adults aged $65+$. We found that while $81.82\%$ of our participants are aware of security features like \lq\lq two-factor authentication (2FA)\rq\rq~ and encryption, $63.64\%$ express serious concerns about unauthorized access to sensitive health data. Only $13.64\%$ feel confident in existing protections, citing confusion over \lq\lq data sharing policies\rq\rq~ and frustration with \lq\lq complex security settings\rq\rq~ which lead to distrust and anxiety. To cope, our participants adopt various strategies, such as relying on family or professional support and limiting feature usage leading to disengagement. Thus, we recommend \lq\lq adaptive security mechanisms,\rq\rq~ simplified interfaces, and real-time transparency notifications to foster trust and ensure \lq\lq privacy and security by design\rq\rq~ in IoT health systems for older adults.
\end{abstract}

\ccsdesc[500]{Security and privacy~Usability in security and privacy}

\ccsdesc[500]{Human-centered computing~Empirical studies in accessibility}

\ccsdesc[500]{Social and professional topics~Seniors}

\keywords{Older Adults, IoT, Security, Privacy, Trust, User Study}


\maketitle

\section{Introduction}
Our daily lives have altered significantly as a result of the Internet of Things (IoT) explosive growth, which has brought us unprecedented interaction, efficiency, and ease ~\cite{Kumar_2019_IoT, NIZETIC_2020_IoT, Salunke_2024_IoT}. IoT technology, which includes wearable health monitoring and smart home systems, promises significant improvements in quality of life for users across demographics ~\cite{Saka_2024_EvaluatingPM, Chataut_2023_IoT, Tazi_Telehealth_2024, Streiff_IoT_2022}. 
In this study, we define IoT devices as network-connected devices that collect, process, and exchange data to support health monitoring and independent living for older adults ~\cite{Chataut_2023_IoT, Liu_2016_SmartHome, Kumar_2019_IoT}. These include wearable such as fitness trackers and glucose monitors that offer real-time health information ~\cite{baig2019systematic, lee2020discrepancies, Saka_2024_EvaluatingPM}, as well as smart home systems like motion sensors and environmental monitors, which indirectly improve health outcomes by enhancing safety and identifying behavioral patterns ~\cite{gonccalves_2021_flexpersonas, Jacobsson_2016_Risk}. This study emphasizes wearables for their direct involvement in health monitoring, while also acknowledging the supplementary role of smart home devices and their implications. Both categories serve complementary roles in facilitating health management and independent living for older adults.
For older adults, these technologies offer critical support for independent living, health management, and aging in place, allowing them to maintain autonomy while managing chronic conditions or daily health needs~\cite{Moraitou_SmartHm, Majumder_2017_Elderly, Liu_2016_SmartHome, Rejeb_2023_IoTHealth}. 
Despite these benefits, significant security and privacy concerns persist, particularly for older adults who may possess limited exposure to training in digital technologies, potentially hindering their ability to understand and manage IoT systems~\cite{CHAURASIA_2023_IoTchallenges, Mazhar_2023_IoT, Virat_2018_IoTC, Tawalbeh_2020_IoTC}. Common vulnerabilities associated with IoT have been identified as weak authentication, insufficient encryption, and unauthorized data access ~\cite{Weber_Challenges, Sicari_2015_Privacy, Iwuanyanwu_2023_IoTrISK, SASI_2023_IoTAttacks, wheeler_userperception_2022}. These challenges are exacerbated for older adults, who may face additional cognitive or physical limitations that hinder their ability to interact securely with IoT devices. 
Cognitive changes associated with aging might affect the ability to identify and respond to security threats, while differing degrees of familiarity with digital technologies frequently hinder efficient configuration and management of privacy settings. 
As IoT becomes more integral to supporting independent living, the very technology meant to enhance their lives leaves older adults vulnerable to cyberattacks and privacy breaches due to limited awareness of digital security practices~\cite{PEREZ_2023_IoTIL, Khaemba_2022_Aging}. 

While there has been extensive research on IoT security and privacy concerns in general, the specific needs and challenges of older adults remain underexplored ~\cite{Ziegeldorf_PrivacyIT, AlZyoud_IoT, Jayashree_2021_Privacy}. Some studies have looked at cybersecurity issues for older adults ~\cite{Nicholson_2019_Cyber, Alves_Fraud, Cassandra_2017_Cyber, Mentis_2020_Cyber, Sheil_cybersecurity_2024} yet few have dig into how these concerns manifest in IoT use, particularly for health monitoring systems.
In contrast to younger users, who may perceive privacy breaches mainly as technical problems, older adults may endure significant emotional distress when their health data is compromised ~\cite{Frik_2019_PrivSec, Morrison_2023_Cyber, Tazi_Priv_Sec_2024}. This intensified reaction arises from various factors specific to this demographic: their generational perspectives on medical privacy ~\cite{Cassandra_2017_Cyber}, heightened reliance on health monitoring for sustaining independence ~\cite{Liu_2024_Smarthom}, and increased consequences of health data breaches due to age-related discrimination ~\cite{Alves_Fraud}. 
Existing solutions designed for the general public fail to address the unique challenges faced by older adults, such as the difficulty in configuring devices, understanding complex privacy settings, and recognizing security threats. Furthermore, the psychological impact of security and privacy breaches, such as anxiety and loss of trust in health monitoring systems, is an area that remains underexplored.
This gap in the literature highlights the need for research that addresses the specific security and privacy concerns older adults face when using IoT devices. 
Our study examines the various security and privacy issues faced by older adults using IoT devices for health monitoring. By examining the barriers encountered, the psychological effects of privacy breaches, and the coping strategies utilized, this work provides a comprehensive understanding of how IoT design can better meet the needs of older adults.
Finally, we offer recommendations for improving IoT security features to better meet the needs of this demographic, with the goal of fostering trust and re-engagement.

Building on these issues, we investigate the perceptions and experiences with respect to IoT security and privacy from older adults in this study. Using a two-phased study (online survey and semi-structured interviews), we collect data. In doing so, we seek to present a comprehensive view of how older adults interact with IoT in terms of their security risks awareness, the challenges they experience when trying to manage those risks and also the psychological impacts associated with potential breaches. This study aims to answer the following research questions (\textit{RQs}):

\begin{itemize}
\item \textit{RQ1: How do older adults perceive the security and privacy features of IoT, and what factors influence their trust and willingness to adopt these technologies?}
\item \textit{RQ2: What barriers do older adults face when configuring and managing the security and privacy settings of IoT?}
\item \textit{RQ3: What impact do security and privacy breaches have on the well-being of older adults using IoT?}
\item \textit{RQ4: How do older adults cope with and seek support for managing the security and privacy concerns of health monitoring IoT devices?}
\end{itemize}

Our research builds upon the existing gaps by offering a user-centered approach, combining both survey and interview data to gain deep insights into the specific security and privacy concerns older adults encounter when using IoT devices. 
Through this study, our work  contributes to the fields of IoT security and privacy for older adults:

\begin{itemize}
\item {Unlike prior studies that focus broadly on IoT users, this work zeroes in on older adults, a group with unique cognitive and physical challenges that can exacerbate security and privacy risks. For example, our findings shed light on how low digital literacy and age-related cognitive decline make older adults more vulnerable to cyber threats, an area that has been underexplored in the literature.}

\item {This study also provides insights into cognitive and emotional barriers, such as paranoia, mistrust, and the high level of frustration associated with handling complex security settings for IoT devices (cognitive overload). These perspectives offer new dimensions in which IoT security challenges can specifically impact older adults as they relate to their mental health and well-being.}

\item {We provide recommendations based on the evidence for designing IoT devices to accommodate older adults. Additionally, our work, suggests usable security by design strategies to reduce barriers to security settings for older users whilst promoting privacy policies that encourage transparency in data handling practice. Thus, these suggestions provide practical direction for IoT developers as well as policymakers and healthcare professionals to strengthen the security-related needs of this demographic.}

\item {Another contribution of this study is the identification of how older adults manage security issues. They typically depend on family or professional assistance or restrict their usage of IoT features because of concerns about privacy and complexity. This finding emphasizes the necessity of producing tailored education and implementing security settings that are more intuitive and adapted to the requirements of older adults.}

\end{itemize}

\section{Related Work}

The intersection of the IoT, security, privacy, and the needs of older adults has gained increasing attention in recent years, especially as IoT devices help healthcare and independent living more and more~\cite{Saka_2023_IoT, Ellefsen_2023_Privacy,baig2019systematic,nathan2018survey,lee2020discrepancies}. Prior studies indicate that demographics like older adults have low trust in new technologies, and are more concerned about data breaches and exploitation which in turn affects their adoption rate~\cite{Schuster_2022_UsersPA, Pakianathan_2020_TowardsID, Mentis_2019_Risk}. 
This section explores prior research through key themes including the barriers older adults face in adopting IoT technologies, the technical vulnerabilities of IoT systems and their implications, the specific cybersecurity challenges those older adults encounter, the psychological and emotional impacts of privacy breaches in health monitoring systems and the ethical considerations necessary for developing more secure and accessible IoT solutions for older adults ~\cite{Tazi_Accessibility_2023, fatima_olderadultsperceptions_2024}. These themes together underscore the necessity for user-centered design and policy interventions to tackle the issues encountered by older adults.

\subsection{Barriers to IoT adoption}
Studies indicate a complex interaction between older adults and the uptake of technology. Although the use of IoT among older adults is increasing, Heart and Kalderon ~\cite{HEART_2013_IoTAdoption} discovered that health-related concerns frequently motivate this adoption more than technological interest ~\cite{Saka_2024_EvaluatingPM}.
Yusif et al.'s work mention that older adults often hesitate to adopt new technologies unless they perceive a clear and direct benefit~\cite{YUSIF_2016_adoption}. Similarly, Jacobsson et al. reviewed smart home technologies for aging in place, highlighting the potential benefits of IoT devices in supporting independent living while noting that security and privacy concerns often act as barriers to adoption ~\cite{Jacobsson_2016_Risk}. Building on this, Lee et al.~\cite{Lee_adoption} and Heart et al.~\cite{HEART_2013_IoTAdoption} explored the factors influencing IoT technology acceptance among older adults, identifying ease of use, perceived usefulness, and trust as key determinants. These studies highlight the necessity of developing useful IoT systems that handle trust-related concerns. Concern over security vulnerabilities in IoT devices, especially those used by older adults, is growing~\cite{alkhatib2018privacy,malhotra2021internet, gonccalves_2021_flexpersonas, ambe_2019_older}.

\subsection{Technical Vulnerabilities of IoT Systems}
In the broader context of IoT security, studies like that of Pourrahmani et al.~\cite{POURRAHMANI_2023_vulnerabilities} and Mazhar et al.~\cite{Mazhar_2023_IoT} have explored the technical risks and identified common vulnerabilities in IoT devices, including weak authentication mechanisms, malware attacks, injection attacks and insecure data transmission. 
Some works have dealt directly with security challenges in smart homes, which also become popular among older adults. In smart home ecosystems, software vulnerabilities, privacy intrusion, man-in-the-middle attacks and denial of service (DoS) were identified as major threats~\cite{Lin_IoTChall, Hall_2020_Smarthome, Liu_2024_Smarthom, BUILGIL_2023_Harm,chhetri2021identifying,anthi2022detecting}. Such software vulnerabilities might let intruders operate IoT devices, threatening older adults' privacy and physical security. 
MitM attacks occur when adversaries intercept data transmitted between IoT devices and a cloud service. The integrity and confidentiality of a communication channel can be compromised if a MitM attacker deceives an IoT device into connecting to their network instead of the legitimate one, thereby capturing sensitive health data ~\cite{Mallik_2019_MiTM}.
Similarly, DoS attacks overwhelm IoT devices with excessive data requests ~\cite{Chao-yang_2011_DoS, Toutstop_Dos_2021}, potentially disrupting essential health monitoring. This could prevent a fall detection system from sending alerts or impede the effective functioning of prescription reminders.
However, their study did not specifically address the implications for older adults, especially in healthcare contexts. 

\subsection{Age-Specific Cybersecurity Challenges}
Prior works on cybersecurity issues facing older adults have grown in recent years, though not always with a specific focus on IoT~\cite{Mentis_2020_Cyber, Zheng_2018_UserPO, Obaidat_2020_ACA, blackwood_2021_cybersecurity, sivagumaran_2023_challenges, morrison_2020_technological}. In studies by Nicholson et al.~\cite{Nicholson_2019_Cyber} and Pacheco et al.~\cite{Pacheco_2024_OlderAS}, they explored older adults' attitudes towards cybersecurity and online privacy, finding that while they are generally aware of cybersecurity risks, they often feel overwhelmed by the complexity of security measures. Similarly, Frik et al. investigated how older adults' privacy attitudes and needs differ from those of younger adults, suggesting that older adults tend to be more concerned about privacy but may lack the technical knowledge to protect themselves ~\cite{Frik_2019_PrivSec}. 
Furthermore, studies by Morrison et al.~\cite{Morrison_2023_Cyber} and Oliveira et al.~\cite{oliveira_2017_security} noted that older adults are frequently victims of social engineering-based cyber-attacks like phishing or impersonation because of their varying levels of familiarity with digital technologies and security practices. These attacks prey on their trust and make it easier to gain unauthorized access to personal data. The risks associated with inadequate security measures rise as IoT devices are increasingly incorporated into health monitoring and smart home environments~\cite{chen2024trustmark}, leaving older adults especially open to misuse~\cite{caven2024comparing,mujirishvili2024don, heinz_2013_perceptions}.
User-centred design approaches have been put forward as a way to create IoT systems that are more secure and reliable for older adults. According to Moreno et al., including older adults in the design process helps boost the technology's acceptance, build trust, and improve usability ~\cite{Moreno_2024_review}. There is an emphasis on designing user interfaces and security settings that require a minimum of technical skill to understand, something highly important for older adults who might struggle with complex configurations. Furthermore, developers can make security features easier to use, like simpler authentication methods or automatic updates by specifically gathering feedback from older adults during the design phase~\cite{Sumner_2020_CoDesign, Darley_2022_Design, Das_designsecure_2024}. Tsai et al. also underscored the need for age-related guides and step-by-step help instructions so that older adults can learn about how to protect their personal information without giving up~\cite{Tsai_OlderAdult}. Though Knowles and Hanson caution that while design technology for older adults has demonstrated its worth, further research is necessary to validate these approaches and explore how they may be used to improve the security and privacy of IoT for older adults~\cite{Knowles_2018_Tech}.

\subsection{Psychological and Emotional Impacts}
An important yet inadequately examined domain in IoT security and privacy research is the psychological and emotional impact of breaches. Previous studies have recorded certain psychological effects when older adults encounter technological disruptions. Morrison et al. discovered that security issues frequently create anxiety regarding technology utilization among older adults ~\cite{Morrison_2023_Cyber}. Anxiety is especially heightened in health monitoring scenarios. Mentis et al. found that privacy violations could trigger strong emotional reactions owing to the sensitive nature of health information ~\cite{Mentis_2020_Cyber}. The psychological consequences appear pronounced for this demographic, as Knowles and Hanson found that older adults exhibit more intense emotional responses to security breaches compared to younger people ~\cite{Knowles_2018_Distrust}, while frequently perceiving themselves as less capable of safeguarding themselves. This emotional burden impacts their mental well-being and affects their willingness to embrace or trust IoT devices. Our work aims to further investigate these psychological effects, particularly focusing on their manifestation in the interactions of older adults with health monitoring IoT devices.

\subsection{Ethical Considerations of Age-Specific Research}
In addition, when talking about IoT security and privacy for older adults, ethical considerations are also crucial. Ethical issues including preserving the privacy and autonomy of older adults continue to be a major factor in the implementation of IoT especially in healthcare environments where sensitive data is involved as highlighted by Friedman et al.~\cite{Friedman_2022_ethic}. The ethical implications of IoT use in elder care have been studied by Finco et al., who have highlighted the necessity for a balance between privacy, safety, and autonomy. For example, although IoT technologies like health monitoring systems and fall detection sensors might improve safety, they frequently need constant data collection, which some users may find intrusive~\cite{Finco_2024_Ethic}. Which, in turn raises ethical questions about what data older adults should have to give up for safety and convenience. It has been suggested that more stringent regulatory frameworks and clear policies are required to prevent the exploitation of older adults and guarantee that IoT devices are safe and privacy-respecting by design~\cite{birchley2017smart,pirzada2022ethics,mcneill2017privacy,mok2015too}.

While existing literature provides insights into the adoption of IoT technologies by older adults, there remains a significant gap in understanding how security challenges and privacy concerns uniquely impact this demographic. Specifically, the intersection of cognitive limitations, digital literacy, and the psychological effects of security vulnerabilities such as anxiety and loss of trust required further study. Additionally, the next step to the prior works that identified some risks is how older adults perceive and respond to these challenges, particularly in health monitoring contexts where the emotional impact might be significant. Also, the coping strategies employed by older adults, such as depending on family assistance or disengaging from IoT, remain underexplored.
Our study addresses these gaps through a two-phased approach, combining survey data and interview narratives to provide insights into older adults' practical and emotional challenges. By focusing on their specific needs, we aim to contribute to more inclusive, user-centered IoT security solutions that promote trust and engagement.

\section{Method}

In this study, we will analyze the security and privacy of IoT from the perspective of older adults. 
The following subsections outline the detailed methodology, including study design, participant recruitment, data collection, data analysis, and ethical considerations.

\subsection{Study Design}
The study utilizes a combination of an online survey and semi-structured interviews to gather data. The online survey, deployed on the Qualtrics platform, consisted of $25$ IoT security and privacy-focused questions, $3$ pre-screening questions, and $3$ demographic questions. The survey included both open-ended and closed-ended questions, allowing for efficient data collection and richer qualitative insights. Attention check questions were incorporated to ensure data quality and for freedom of expression and comfortability, we made all questions optional except the pre-screening questions. We conducted semi-structured interviews with a subset of participants from the survey phase who volunteered to take part. These interviews, lasting about $20-30$ minutes, consisted of $18$ open-ended questions designed to explore participants' experiences, concerns, and perceptions in greater depth. Participants were asked conventional demographic questions such as age, education, gender and ethnicity. For compensation, there was no compensation for those who took part in the survey but participants who went ahead to take part in the interview received a $\$10$ Amazon gift card as a token of appreciation.

Our target population were older adults aged $65$ and above who either use or intend to use IoT devices in their daily lives. We employed multiple recruitment channels, including email outreach through aging institutes, flyers, social media posts, and word-of-mouth. Interested individuals completed a pre-screening questionnaire to assess their eligibility based on age, U.S. residency and honesty in answering the questions. The purpose of the study, its procedures, and the importance of the participant's participation in contributing to enhancing IoT security and privacy for their demography were made clear to the participants. For the online survey, we initially recruited $29$ participants, with a final sample size of $22$ after excluding incomplete responses and those who failed attention checks. Qualtrics XM survey platform was used to collect the survey data and participants completed the survey at their own pace. We conducted interviews with $9$ participants who agreed to do so through phone calls and in-person appearances. With participant consent, interviews were audio-recorded for later transcription and analysis.

\subsubsection{Pilot Study}

Before we began the data collection, we conducted a pilot study with $10$ random participants. The purpose of this pilot was to allow us to improve our questionnaires and make sure they adequately addressed our study concerns. Participants in the pilot study were asked to complete the survey and provide feedback on the clarity of questions, ease of use of the survey platform, and overall survey experience. For the interview, we conducted mock interviews to test the flow of questions and identify any potential issues. We improved our survey and interview approaches by making the required changes in response to the feedback we got during this process.

\subsubsection{Survey Design and Interviews}
At the start of both the survey and interviews, we provided participants with a definition of IoT in the context of our study: ~\lq\lq Internet of Things (IoT) devices, also known as Smart Devices for older adults, are technological devices equipped with sensors, connectivity, and data processing capabilities that enable data collection and data exchange over the internet. These devices are specifically designed to cater to the needs of older adults, providing them with enhanced convenience, safety, and assistance in various aspects of daily life \rq\rq~.
The first phase, the online survey, was deployed on the Qualtrics platform, chosen for its popularity and user-friendly interface suitable for older adults.
The survey was developed by adapting the Senior Technology Acceptance Model (STAM) ~\cite{Chen_2014_GerontechnologyAB}. These frameworks specifically address older adults' technology acceptance patterns and concerns.
The survey began by explaining the study's purpose, procedure, and consent requirements. After collecting participants' informed consent, they answered the pre-screening questions to ensure they met the eligibility criteria of being $65$ years or older, residing in the United States, and willing to give honest answers. Those who failed to meet these criteria were not allowed to continue with the survey. 
The survey was made up of $25$ questions which were organised into four thematic areas and $3$ demographic questions. These areas included IoT Usage and Adoption, Security and Privacy Awareness, Risk Perceptions and Concerns, and Coping Strategies and Support.
Closed-ended questions were used to collect data on older adults' views and experiences with IoT. Open-ended questions were included to allow participants to express their thoughts freely, providing detailed qualitative data. All questions except the pre-screening ones were made optional to ensure participant comfort. Attention check questions were placed within the survey to ensure participants were focused, and responses from participants who failed these checks were excluded from the final analysis. Finally, participants were asked demographic questions and any additional comments they had. The complete survey questionnaire is included in ~\autoref{app:survey}.

For the second phase, we conducted semi-structured interviews with a subset of participants to complement the survey data. 
This phase was created to expand upon the findings obtained in Phase one. The initial analysis of survey data identified significant areas of concern, including the emotional impact of security breaches, challenges managing privacy settings, and support networks for technology utilization. These ideas guided the formulation of interview questions, enabling an exploration of participants' emotional response, coping strategies, and comprehensive experiences with IoT devices.
These interviews, lasting about $20-30$ minutes, were designed to explore older adults' experiences, concerns, and perceptions in detail. The interview consisted of $18$ open-ended questions covering topics such as personal experiences with IoT devices, perceived benefits and risks, challenges in managing IoT security and privacy, sources of information about IoT security, and suggestions for improvement. Those who engaged in the interview session were rewarded with a $\$10$ Amazon gift card. The decision to participate in the interview was entirely voluntary. Participants indicated their interest at the end of the survey by providing their email addresses for contact regarding the interview phase. The survey was completed by $29$ participants, but only $9$ of them showed interest in taking part in the interview and were subsequently invited to participate. The semi-structured format allowed for flexibility in the conversation, enabling us to probe deeper into the interesting responses of participants. The interview questions are included in ~\autoref{app:interview}.

\subsection{Ethical Considerations}
This study was approved by the University's Ethical Review Board, with specific attention to protecting older adults. We prioritized the protection of participants' privacy and well-being throughout the research process. Informed consent was obtained from all participants, with survey completion considered implied consent and verbal consent recorded for interviews.
Due to the delicate nature of discussing privacy and security concerns with older adults, we implemented comprehensive ethical safeguards. The pre-interview briefings explained participants' right to skip questions they find uncomfortable or to interrupt conversations, and a system for addressing emotional distress was implemented, including the option to terminate the interview at any moment. Before commencing our research, we the researchers undertook courses pertinent to emotions to recognize signs of discomfort during interviews, and we did regular check-ins with participants who recounted potentially distressing situations. All interviews were carried out with careful attention to participants' emotional welfare, including provisions for breaks and transcript evaluation.
Data handling procedures prioritized participant privacy, with careful anonymization of health-related information. All collected data was anonymized and stored securely on encrypted drives, accessible only to the researchers of this study. Given our focus on older adults, we took extra care to ensure that our materials and procedures were accessible.

\subsection{Analysis Strategies}
The data set obtained by both the survey and interview phase means are integrated to offer comprehensive insights into IoT security as well as privacy perceptions among older adults. Descriptive statistics were used to summarise participant demographics and the distribution of responses from our survey, regarding levels of awareness, concern and perceived effectiveness of IoT security features. 
Inferential statistical analysis was not conducted because of our study's exploratory nature and limited sample size; however, the descriptive findings yielded significant insights into trends and patterns.
Thematic analysis was used to analyze qualitative data obtained from the interviews, to recognize common themes and trends across the issues that older adults encountered, as well as their opinions about IoT security and privacy. All interviews were recorded and transcribed using \lq trint\rq~ commercial processing tool ~\cite{trintTranscriptionSoftware} and were all manually verified by the first author.
Two researchers independently coded the interview transcripts, resulting in an inter-coder reliability coefficient of 0.84 (Cohen's kappa). Preliminary codes were categorized into themes via iterative team discussions, focusing on persistent trends in security issues, privacy perceptions, and emotional responses to technology use. The coding procedure concentrated on identifying explicit and implicit references to the psychological effects of security breaches, barriers to technology adoption, and coping strategies. The final themes were evaluated and refined to ensure they reflected the participants' experiences and perspectives.
These qualitative descriptions were subsequently combined into the larger narrative alongside our survey data, allowing us to create a more detailed and contextualized understanding of it all, guaranteeing that the study's conclusions were supported by facts and accurately represented the experiences of the participants.

\section{Result}

Our research used a two-phased study, combining survey data with qualitative insights from semi-structured interviews. This section presents the findings from both components.

\subsection{Survey -- Phase-I}

Our online survey gathered responses from $29$ participants administered through the Qualtrics platform, with $22$ valid responses after excluding incomplete submissions and those that failed attention checks. The survey presents the findings of older adults' perceptions, behaviors, and concerns regarding IoT device security and privacy. Our respondents were 65 years of age or older, aligning with our target demographic of older adults. Gender distribution was relatively balanced, with $55\%$ identifying as male and $45\%$ as female. Participants were mostly white ($59\%$), with some representation from Latino or Hispanic backgrounds ($14\%$) and those of African Origin ($18\%$). Educational backgrounds varied, the majority of participants reported having at least some college education, with the largest group ($45\%$) holding master's degrees, followed by bachelor's degrees ($23\%$), and vocational training ($9\%$). These demographic details are summarized in ~\autoref{tab:Demographic}, providing an overview of the participants' backgrounds.

\begin{table}[!ht]
\centering
\caption{Demographic Characteristics of our Survey Particiapnts}
\begin{tabular}{|l|l|c|}
\hline
\textbf{Characteristic} & \textbf{Category}          & \textbf{Percentage} \\ \hline
Gender                  & Men                       & 54.55\%             \\ 
                        & Women                     & 45.45\%             \\ \hline

Education               & Masters degree             & 45.45\%             \\ 
                        & Bachelors degree           & 22.73\%             \\ 
                        & Vocational Training        & 9.09\%             \\ 
                        & Other                      & 22.73\%             \\ \hline
Ethnicity               & White                      & 59.09\%
\\ 
                        & Black/African-American         & 18.18\%
\\ 
                        & Latino or Hispanic         & 13.64\%             \\ 
                        & Others                & 9.09\%              \\ \hline
\end{tabular}
\label{tab:Demographic}
\end{table}

\subsubsection{IoT Device Usage and Awareness}

\begin{figure}[htbp!]
\centering
\includegraphics[width=1\linewidth]{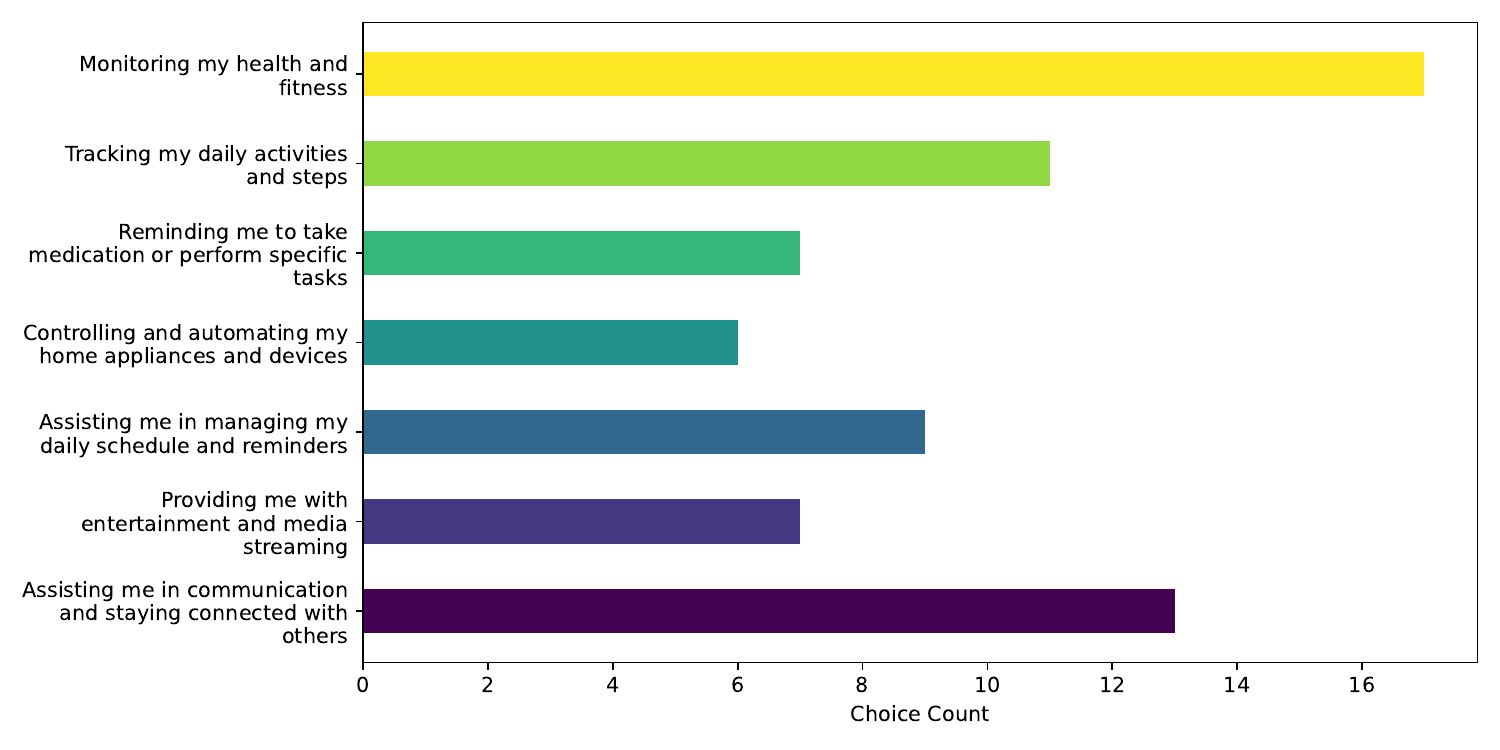}
\caption{How do you typically utilize your IoT device in your daily life?}
\Description{Horizontal bar chart showing IoT device usage patterns among older adults. The chart displays seven categories of use, with choice counts. Health monitoring and fitness ranks highest, followed by Assisting in communication and staying connected with others (13 counts). Tracking daily activities and steps (11 counts), Assisting in managing daily schedules and reminders (9 counts), while medication reminder and entertainment both have 7 counts. Controlling and automating home appliances and devices has the least count. Data presents clear prioritization of health-related functions over entertainment features. The x-axis shows choice count and y-axis lists usage categories}
\label{usage}
\end{figure}

We gathered information from our survey about participants' usage patterns and understanding of security and privacy aspects, as well as their experiences with and concerns about IoT devices.  Most had used IoT or similar wearable devices for health monitoring and activity tracking in the past year ($91\%$), indicating a high rate of adoption in this population. The most common purposes the participants reported for using these devices were monitoring health and fitness ($24.29\%$), communication and connection with others ($18.57\%$), daily activities tracking and steps counting ($15.71\%$), and managing schedules and reminders ($12.86\%$). This significant presence of IoT-based smart devices in the daily living environment highlights their emerging role in supporting different aspects of well-being and independence among older adults as the choice count is shown in ~\autoref{usage}. Our research shows that while IoT is very common, there is still a concerning gap in translating awareness of security and privacy features into action. On the positive side, $81.82\%$ of respondents confirmed knowledge about security functions like two-factor authentication (2FA), data encryption, and privacy protection but practice with these components was abysmally low as well. For example, $63.64\%$ of our respondents rarely reviewed or updated the privacy and security settings on their IoT devices, with an additional $13.64\%$ never doing so. The gap between understanding and keeping their information safe digitally highlights a critical area for improvement in user education and interface design.

\subsubsection{Security and Privacy Concerns}

\begin{figure}[htbp!]
\centering
\includegraphics[width=1\linewidth]{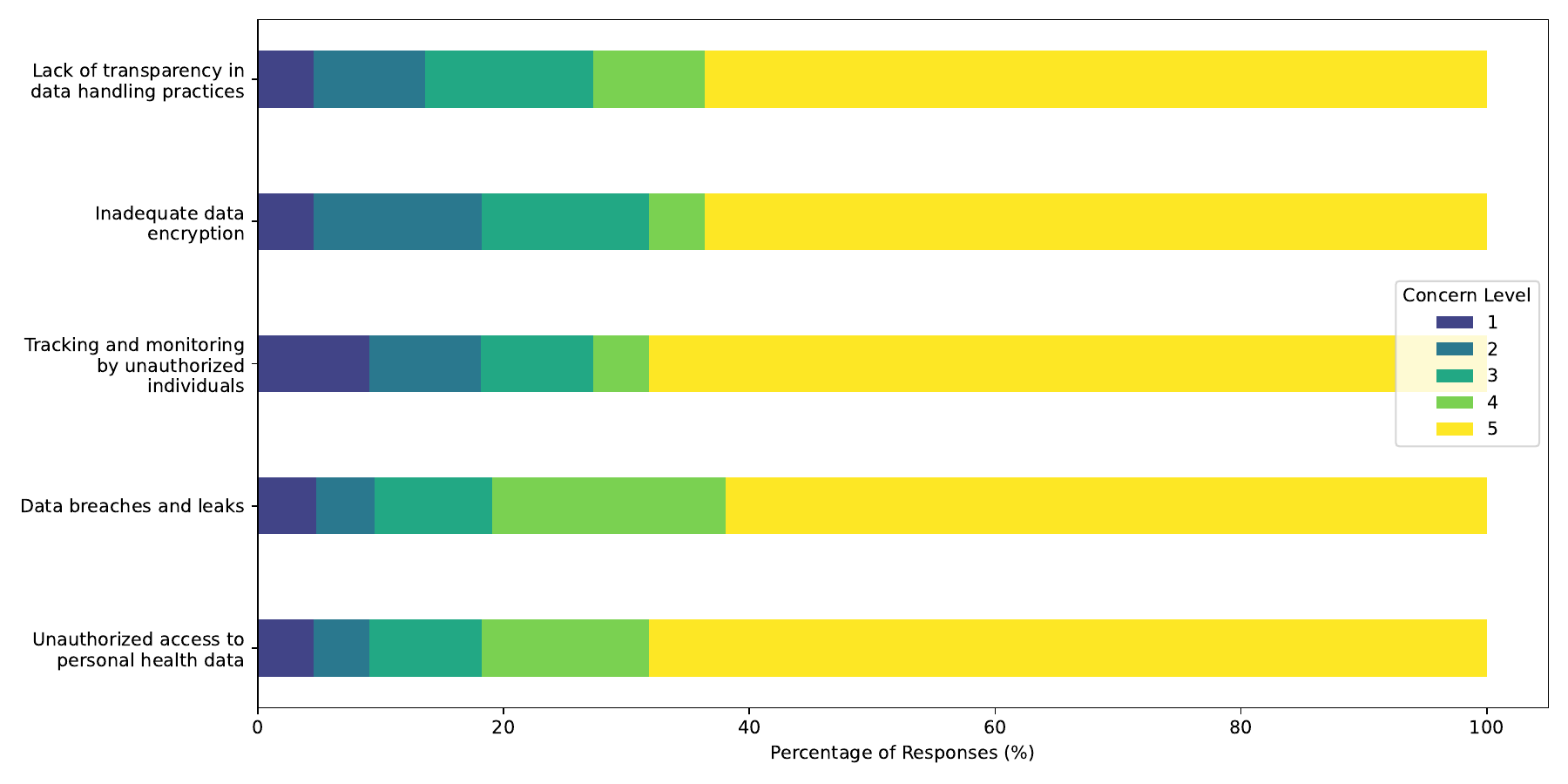}
\caption{Please rate the following security and privacy concerns regarding IoTs on a scale of 1 to 5, where 1 represents "Not Concerned" and 5 represents "Very Concerned"}
\Description{Stacked horizontal bar chart depicting privacy and security concerns about IoT devices on a scale of 1-5, where 1 represents 'Not Concerned' (dark blue) and 5 represents 'Very Concerned' (yellow). The chart shows five areas of concern: unauthorized access to health data, data breaches, unauthorized tracking, inadequate encryption, and lack of transparency. For each category, the percentage distribution of responses is shown from 0-100\%. Around 60-70\% of responses indicate high concern levels (4-5) across all categories, with unauthorized access to personal health data showing the highest proportion of Concerned responses.}
\label{concern}
\end{figure}

While we dug deeper into issues around security and privacy surrounding IoT, our study found a strong level of concern among older adults about data safety while using an IoT device with $64\%$ of participants being \lq\lq Very concerned\rq\rq, and $14\%$ were \lq\lq Somewhat concerned\rq\rq~. When asked to rank their level of concern on a range of topics, from $1$ (Not Concerned) to $5$ (Very Concerned), regarding things like data breaches, unauthorized access to personal health information, and inadequate data encryption. "Unauthorised access to personal health data" was rated as one of the most significant ($\mu$ of $4.36$ out of $5$) followed by data breaches and leaks ($4.29$), and lack of transparency in data handling practices ($4.18$). ~\autoref{concern} summarizes participants' responses with high ratings in a variety of areas of concern implying that older adults are aware of the possible risks associated with these IoT devices. It is interesting to notice that the participants were worried about their security and privacy even though their confidence in IoT security solutions varied. When asked about the effectiveness of IoT's security and privacy measures, $14\%$ considered them very effective, $55\%$ somewhat effective, $18\%$ were neutral, and $9\%$ considered them not effective. This divergent distribution indicates that older adults and IoT technology may have a nuanced relationship, in which an awareness of the potential benefits coexists with continuing concerns about security adequacy.

\subsubsection{Challenges and Impact of Security Incidents}

Results from our survey indicate some of the challenges older adults face with securing their IoT data, and what effect experiencing a security incident may have. Complexities associated with IoT Technology emerged as a major constraint, when asked about the barriers to security measures some participants' response was difficulty in deducing what security settings should be and how to implement them. As one respondent put it: \lq\lq Difficult understanding what the settings and security should be and what they mean,\rq\rq~ which speaks to more user-friendly interfaces (a common complaint) alongside clearer communication around basic security features.
Technical literacy was another of the common challenge. Those who did use this excuse, also stated \lq\lq lack of computer literacy\rq\rq, \lq\lq understanding the internet language\rq\rq, and \lq\lq confusing new technology\rq\rq~ as obstacles to properly managing IoT device security. This technology gap often means that older adults are exposed not only to security risks of which they may be unaware but also ones that are impossible for them to mitigate. There were other challenges in terms of the physical aspects of device management. A participant also noted that \lq\lq too often the instructions are hard to understand, the print is too small and it's color\rq\rq.  This points out the importance of accounting for physical concerns when designing IoT for this population.

Regarding security incidents, more than three in five ($63.64\%$) noted that they're \lq\lq very concerned\rq\rq~ about the security and privacy of their personal information when using IoT devices while less than half of the participants have experienced any type of incident thus far. Over half of the participants ($68.18\%$) said they experienced no security or privacy incidents with IoT or similar devices when asked about any that occurred. However, among those who reported having an incident, there was a variety of problems. These manifestations ranged from stolen funds in bank or credit card accounts, use of personal information to commit financial fraud and exposure to scams. One participant reported, \lq\lq Someone getting access to my bank account. Paranoia and mistrust were the result,\rq\rq~ highlighting the psychological impact of such breaches. Another respondent reported, \lq\lq Medicare notice re: mine has been hacked\rq\rq~ as an example that not even systems which seem to concern healthcare can be said to be safe from harm. Even more surprisingly, despite the challenges and incidents, $82\%$ of respondents selected either \lq\lq Very Likely\rq\rq~ ($36\%$) or \lq\lq Likely\rq\rq~ ($46\%$) about recommending IoT for other older adults. In other words, there are security issues but some older adults also see the value in IoT and believe that the benefits outweigh the risks. The convergence of these challenges results in the exposure of older adults to IoT security risk.

\subsubsection{Sources of Information and Recommendations}

The older adults in our survey seek information on IoT security and privacy from several different sources. The primary source was official IoT communications, cited by $45.45\%$ of participants, followed by news articles and blogs ($31.82\%$), indicating a preference for traditional media. Also, relatively few participants opted to use social media ($9.09\%$) or online forums ($4.55\%$), presumably because these individuals were trustful of them. Family members were also cited as a source of information, although this only applied to $9.09\%$. In addition, participants shared helpful thoughts about how to make the IoT more secure. Some requested more transparency and easier configuration of security settings, with one respondent saying that he needed \lq\lq more transparent controls on what data is shared and confidence that my choices are respected\rq\rq. There was a similar desire for more effective notifications of personal information use and better user education targeted at older adults. A few of the recommendations made were that fingerprint authentication be used for security purposes, automatic updates are delivered to all software products and finally easy-to-follow methodological steps on how an end-user can take control over their Security settings. Concerns about data privacy were also raised with the call for more information about data aggregation and Personally Identifiable Information protection.

\subsection{Semi-Structured Interviews -- Phase-II}

To complement our survey findings, we conducted in-depth interviews with nine participants ($P1$ – $P9$) as part of our study on the security and privacy of IoT among older adults. For the nine people who participated in the interviews, we use pseudonyms ($P1$ – $P9$) to protect their privacy and confidentiality. This allows us to explicitly reference their comments throughout the study while maintaining the anonymity of their identities. The interviews provided in-depth insights into participants' experiences, concerns, and expectations regarding IoT devices used for health monitoring and independent living. The results are organized into key themes that emerged during the analysis, each supported by quotes from the participants to illustrate their perspectives.

\subsubsection{Adoption and Motivation for IoT Use}

The health conditions as well as the wish to remain self-sufficient influenced individual experiences and adoption of IoT devices. The reasons for adoption varied, but health monitoring emerged as one of the main drivers, with participants highlighting diverse uses of IoT technology to support their well-being and independence. 
For instance, $P1$ utilized sleep monitoring technology including a sleep pad device and AirSense CPAP machine, while $P2$ relied on diabetes management devices like the Libre and Dexcom insulin pumps. 
For example, $P9$ shared that they use a medication reminder and a blood sugar monitor regularly, which have become essential in managing their diabetes: \begin{quote} \lq\lq My health is my primary motivator; without these devices, managing my condition would be much more difficult. The medication reminder helps ensure that I don't miss my doses, which is something I struggled with before.\rq\rq  \end{quote} This example shows how IoT devices might be very helpful in managing chronic diseases, possibly leading to better health outcomes and lessening the cognitive burden that comes with drug schedules. In the same vein, $P3$ expressed a focus on health monitoring routine, \begin{quote} \lq\lq I use a Fitbit... I've been using it to monitor my steps, monitor my sleeping pattern, that's my experience. I mostly use for monitoring. My heart rate variation, beats per minute, footsteps and sleeping conscience.\rq\rq  \end{quote} $P3$'s experience demonstrates the complex nature of health tracking made possible by IoT devices, which includes vital signs and sleep habits in addition to physical activity. This monitoring may help identify conditions early on by giving older adults a broader overview of their health. Similarly, safety and independence were also key motivators, demonstrated by $P5's$ adoption of the Kanega Watch for fall detection, while P7 used Amazon Echo Dot to overcome mobility and hearing challenges and the benefits of tracking daily activities were also captured by $P4$ saying, \begin{quote} \lq\lq I started using these devices to track my daily activities, monitor my health metrics like heart rate, blood pressure, and receive reminders for my medications.\rq\rq \end{quote} The integration of health monitoring with daily activities is highlighted by $P4$'s comment, which implies that IoT devices can easily include health management in regular activities. However, not all participants had begun using IoT devices. $P8$, in particular had potential interest surrounding using IoT devices in the future however remained hesitant due to privacy concerns: \begin{quote} \lq\lq I'm starting to realize their importance, especially as I'm getting older, but I'm also hesitant because I've heard stories about privacy issues.\rq\rq \end{quote} This draws attention to a significant adoption barrier as worries about data security and privacy may make even older adults who perceive the benefits of IoT devices hesitate to adopt them. These diverse experiences demonstrate that although health monitoring remains a powerful driver of IoT device adoption among older adults, challenges like security and privacy issues continue to impede widespread adoption.

\subsubsection{Security and Privacy Concerns}

Discussions often focus on security and privacy, with people feeling very wary about protecting their personal information. 
Our survey indicated $64\%$ of participants are \lq\lq Very concerned\rq\rq~ about protecting their personal information and unauthorized access to health data rated as their highest concern ($\mu$ of $4.36$ out of $5$). These quantitative concerns were demonstrated in our interviews as many older adults are conflicted between the need to monitor their health and the risks of improper handling of their data. The expression \lq\lq Watch My Health, Not My Data\rq\rq~ perfectly captures this feeling and concern that older adults have when attempting to weigh the benefits of IoT devices against the possible concerns of data exposure.
$P6$ had this to say about it, which isn't surprising considering he experienced identity theft multiple times due to his PayPal being hacked: \begin{quote} \lq\lq I've had my identity stolen a couple of times. Recently this person got in and hacked my account for about $\$75,000$, of which I was able to block before they sent out the money of which I didn't have. So, I don't use PayPal to the extent that I did before because it was a real big concern for me.\rq\rq \end{quote} The incident demonstrated how security breaches may undermine trust and restrict the use of potentially helpful IoT technologies. Our survey data reinforces this, indicating that although $81.82\%$ of participants were aware of security features like two-factor authentication, only $13.64\%$ expressed confidence in utilizing these protections.
Similarly, $P9$ also said they were anxious about who got their health data: \begin{quote} \lq\lq The thought of my health data getting into the wrong hands is very concerning. I'm particularly worried about identity theft or my information being used for scams.\rq\rq \end{quote} This concern is a reflection of anxiety among older adults that the very information intended to help them stay healthier could be used against them, leading to a condition where stress is generated by health monitoring.
$P6$ gave an example from a personal perspective: \begin{quote} \lq\lq I had to change my credit card account, so I had to change my banking account with the credit union. I had to change all of the automatic payments that I have now... that seems extremely invasive to me.\rq\rq \end{quote} This shows how security breaches can cause financial stress in addition to affecting daily life and making it more difficult for people to trust IoT devices. Participants also worried about harvesting user information. Concerns about data harvesting were best captured by P8 who stated that: \begin{quote} \lq\lq Data harvesting is a definite concern for me. I've read a lot about how these systems can be hacked, and personal information gets stolen. It's frightening to think about who might have access to your health data… The thought of someone out there having all this information about me without my full knowledge or consent is unsettling.\rq\rq \end{quote} $P8$'s response brings to light the uneasiness that many participants have regarding the invisible nature of data collection and its possible mishandling. These examples highlight the conflict that exists between the need to protect privacy and the need to monitor health conditions. In order to increase older adults' trust and confidence in IoT devices, their concerns regarding data security and privacy must be addressed as these devices become more and more integrated into their daily lives.

\subsubsection{Benefits vs. Risks}

Although security and privacy were also common themes but most of the participants recognized several benefits of using IoT devices to monitor health or for independent living. $P7$ noted, \begin{quote} \lq\lq Right now, I feel the benefits actually outweigh the security and privacy concerns ... I am able to keep my lights off when I need to sleep and I can easily ask my IoT to actually turn on the lights when I need to either use the bathroom in the middle of the night or when my Fitbit prompts me to get either my heart rate up because it's quite low.\rq\rq \end{quote} saying that being able to keep an eye on health measurements in real-time was a great help in maintaining independence. $P9$ expressed this, especially in the context of health management: \begin{quote} \lq\lq I'm using these devices because I care deeply about my health, which, in my view, is more important compared to the risks. Managing my diabetes effectively is my top priority, and these devices play a crucial role in that.\rq\rq~ \end{quote} However, the balance between benefits and risks varied among participants. $P8$ added that as much as he would like to have IoT devices the security and privacy concern takes priority, \begin{quote} "the potential risks related to privacy and security might outweigh the benefits unless I'm convinced that these devices are safe to use". \end{quote} This further underscores the importance of strong security protocols required to obtain trust from older adults.

\subsubsection{Challenges Faced by Older Adults}

Interviews highlighted various issues that older adults face using IoT devices, primarily concerning configuring security settings and navigating privacy options. $P6$ pointed out the difficulty in keeping up with technology: \begin{quote} \lq\lq The biggest issue that we as seniors have is the fact that we don't remember our passwords… The new technology is surpassing the ability of seniors to keep up with it.\rq\rq \end{quote} $P4$ also listed the difficulty of understanding and managing complex security settings as one of their most important challenges: \begin{quote} \lq\lq The most critical challenges include understanding and managing complex security settings. Staying informed about potential threats and ensuring that devices are regularly updated with the latest security patches.\rq\rq \end{quote} Lack of adequate information and support was another important challenge as $P7$ was frustrated with privacy policies he described as confusing pieces of text that are difficult to read and understand: \begin{quote} \lq\lq I feel like if these privacy policies could be made more interactive, that would actually go a long way in allowing users to understand better what they are actually agreeing to, because I get fatigued having to read three, four pages of privacy policies.\rq\rq \end{quote} There was a reference to physical limitations as $P6$ mention that \begin{quote} \lq\lq our biggest issues here are as seniors, we don't see well, nor do we hear well. And probably $30\%$ or more of us here have some form of dementia.\rq\rq \end{quote}

\subsubsection{Trust and Confidence in IoT Companies}

$P6$ in particular, called out skepticism of IoT companies and promises around data security and privacy. P6 stated \begin{quote} \lq\lq I don't feel adequately informed about how my data is collected, stored, and shared. It seems like most of these companies are after profit, and they don't make it easy for users to understand what's happening with their data.\rq\rq \end{quote} It's a feeling that generally extends across participants that not enough IoT companies are keeping user privacy top-of-mind. However, there was also an acknowledgement that some companies are working on improving security. However, that still hasn't been enough to fully address participants' concerns. $P8$, for example, acknowledged that while some security features exist, they still felt uneasy about the potential risks: \begin{quote} \lq\lq I'm aware of some security features, but I'm not sure how effective they really are. Without transparency, it's hard to feel confident in their data handling practices.\rq\rq \end{quote} These diverse viewpoints underscore the complicated relationship older adults have with IoT devices and the companies that produce them.

\subsubsection{Expectations and Recommendations}

Across the board, participants talked about wanting more transparency regarding IoT security and privacy as well as education and easy-to-use options. $P6$ recommends that IoT companies hold seminars or other informational sessions on older adults: \begin{quote} \lq\lq I would like to present a seminar at my facility… to help them figure out what to do if it gets misused.\rq\rq \end{quote}  $P4$ highlighted the fundamental enhancements of streamlined security settings and easier-to-use interfaces: \begin{quote} \lq\lq Simplified security settings, better educational resources, and more intuitive user interfaces will be beneficial.\rq\rq \end{quote} They also stressed the importance of updating and communicating transparently with users about data practices. P8 notes that it should be easier for them to know who has their data and how they are protected: \begin{quote} \lq\lq Companies should make it easier for users to understand how their data is being used and provide clear, straightforward ways to protect it.\rq\rq \end{quote} Again, $P9$ echoed a similar sentiment in that \begin{quote} \lq\lq These companies should be more transparent about their data practices, possibly through regular updates or reports that explain in plain language what's happening with our data.\rq\rq \end{quote}

Interview responses indicate a nuanced relationship that older adults have with IoT devices, in which they acknowledge the potential to accrue substantial benefits from these technologies alongside deep concerns about security and privacy. The results of the research indicate that there is an urgent need for IoT solutions that are secure and more usable to attract older adults. Participants' suggestions for easier security options, improved user training and increased transparency of data practices show important information to guide the design of IoT devices that are more closely aligned with this demographic.

\section{Discussion}

Our two-phase study provides valuable insights into how older adults interact with IoT technologies for health monitoring. Before addressing our research questions, we explore this demographic’s unique challenges with IoT devices, particularly concerning privacy and security. These age-specific challenges, rooted in emotional, physical, and cognitive factors, fundamentally shape their interactions with and adoption of technology.
Our findings corroborate and expand upon prior studies of older adults’ engagement with IoT technologies. For instance, Frik et al. identified that older adults typically exhibit stronger privacy preferences than younger individuals ~\cite{Frik_2019_PrivSec}. Our study reveals how these preferences manifest specifically in the context of health monitoring, with increased concerns about health data breaches aligning with Morrison et al.'s conclusions on the susceptibility of older adults to cyberattacks ~\cite{Morrison_2023_Cyber}. These findings also extend Heart and Kalderon’s ~\cite{HEART_2013_IoTAdoption} research on technology acceptance by highlighting the significant impact of privacy concerns on health technology adoption.
By examining participants’ concerns, the barriers they face, and the impact of security breaches, this work provides an understanding of how older adults interact with IoT technologies and the factors influencing their trust and adoption. In contrast to prior research that broadly examined IoT security, our findings emphasize age-specific vulnerabilities, such as limited digital literacy and usability challenges, that create barriers to efficient security management. These insights underscore the importance of improving awareness and designing more accessible, user-friendly interfaces to support older adults in managing security and privacy settings effectively.

\subsection{Age Specific Considerations}

Our study shows findings unique to older adults that highlight how their interaction with IoT monitoring devices differs from younger populations, particularly in relation to security and privacy concerns. These results, which are specific to older adults, highlight the difficulties that this group encounters because of emotional, physical, and cognitive factors. 

\begin{itemize}

\item {Cognitive and Physical Barriers:}
The most frequent challenge experienced by older adults was the complexity associated with taking charge of IoT security settings. Many participants struggled with understanding technical jargon and navigating confusing interfaces, something not so common among younger IoT users.  As one participant put it, \lq\lq The instructions are hard to understand, and it's not clear what these settings do\rq\rq. Another mentioned memory problems as a reason for an inability to handle IoT security. When asked what the respondent thought was a major problem for older adults, they replied: \lq\lq The biggest issue that we as seniors have is the fact that we don't remember our passwords\rq\rq. This challenge is more pronounced in older populations and directly impacts their ability to maintain secure practices.
These cognitive challenges were further compounded by reduced vision and dexterity faced as one ages, making it even more difficult for older adults to navigate through small buttons on screens in order to set up security features efficiently.

\item {Digital Literacy and Mistrust in Technology:}
One of the common themes amongst our interviews is that older adults have a limited level of digital literacy which raised concerns about cyber security breaches or privacy violations. Older adults are also more sceptical than younger generations, who often trust IoT devices or view the risks to their privacy as a part of using technology ~\cite{Knowles_2018_Distrust}. As one respondent put it: \lq\lq I'm not sure what information is being collected or who sees it\rq\rq, highlighting the extent of distrust in IoT vendors. Such mistrust of these devices, fostered by technological unfamiliarity, provides an even more significant challenge to their integration and the secure use of IoT among older adults.

\item {Emotional Impact of Security Breaches:}
Older adults also expressed stronger emotional responses to potential security and privacy incidents. Some shared paranoia, stress, and anxiety, in discussing this fear of the possibility that their data would be unsafe. $P9$ explained, \lq\lq It's scary to think that something meant to help with my health could potentially be a gateway for someone to steal from me, I don't feel safe\rq\rq. In older populations, the emotional response to breaches is heightened more than in younger ones who might not worry as much about a potential breach. After examining the challenges encountered by older adults, we will now focus on addressing our research questions.

\end{itemize}

\subsection{Perceptions of Security and Privacy}
\textbf{RQ1:} \textit{How do older adults perceive the security and privacy features of IoT, and what factors influence their trust and willingness to adopt these technologies?}\\
Our study reveals a complex and sometimes contradictory relationship between older adults and the security and privacy features of IoT devices such as our studied health monitoring systems. While IoT offers great potential to support health monitoring and independent living, trust in these technologies is often shaky. In our survey, 64\% of participants reported being \lq\lq very concerned\rq\rq~ about the security and privacy of their personal data. This concern wasn't just theoretical; in our interviews, participants expressed very real fears about the misuse of their information. For example, $P9$ said: \lq\lq The thought of my health data getting into the wrong hands is very concerning. I'm particularly worried about identity theft or my information being used for scams.\rq\rq~ These personal concerns show that anxiety about data security continues to prevent older adults from fully trusting IoT devices.

What is particularly surprising is that this fear persists despite a solid awareness of security measures. Our data shows that 81.82\% of respondents are familiar with features like two-factor authentication and data encryption, but only 13.64\% believed these protections were \lq\lq very effective\rq\rq~ at keeping their data safe. This mistrust in the device usage and their security and privacy features is concerning. As $P8$ noted, \lq\lq I know there's encryption, but I don't know if it's really enough to protect my data.\rq\rq~ This highlights a core issue where simply knowing that security features exist doesn't necessarily make older adults feel secure. Often, older adults are stereotyped as lacking cybersecurity knowledge and are labeled as \lq\lq vulnerable\rq\rq~\cite{eilenberger_2022_becoming, park_2017_formal, culo_2011_risk}. However, our data shows that older adults are well aware of security practices, but the constant stream of data breach news, combined with their own negative experiences, leaves them feeling disempowered and unsure whether these security and privacy features are truly effective. Rather than a lack of understanding, it's this erosion of trust that undermines their confidence in IoT devices.
Our findings build on the research of Frik et al., which emphasizes privacy concerns among older adults in contrast to younger populations ~\cite{Frik_2019_PrivSec}. Our study shows how these concerns manifest emotionally, with participants expressing anxiety and distrust especially associated with health monitoring devices. 

Trust, therefore, emerges as the most crucial factor in whether older adults adopt IoT technologies. Participants acknowledged the many benefits of IoT, especially in monitoring their health and maintaining independence, but they remained cautious about fully embracing these devices. This is consistent with previous research showing that trust plays a vital role in the adoption of technology by older adults~\cite{Lee_adoption}. Yet, in our findings, trust is closely tied to transparency. Throughout the interviews, participants frequently expressed frustration over the lack of clear communication from IoT companies about how their data is handled. $P8$ voiced this frustration directly: \lq\lq It's hard to feel confident without knowing exactly how our data is being used.\rq\rq~ This lack of transparency leaves many older adults feeling vulnerable and unsure about who has access to their data and for what purpose. For these participants, it's not enough to have security measures in place; they need clarity and reassurance that their data is being handled with care.

What stands out from our findings is the emotional weight older adults attach to data security. Their concerns are not just abstract fears of potential breaches but real, personal worries about financial safety and well-being. For many participants, it's not just about privacy in a digital sense—it's about the fear that their health data could be exposed or misused, with potentially serious consequences. The emotional burden of these concerns suggests that IoT companies need to do more than just build secure systems. They must communicate that security in a way that feels accessible and trustworthy to older users. If these concerns are not addressed, the promise of IoT for improving the health and independence of older adults may remain unfulfilled. While participants see the benefits of IoT devices, these benefits are overshadowed by ongoing fears about the safety of their personal information. Without the necessary trust in these technologies, older adults are likely to remain hesitant, limiting the potential impact of IoT on their daily lives.

\subsection{Barriers to IoT Security}

\textbf{RQ2:} \textit{What barriers do older adults face when configuring and managing the security and privacy settings of IoT?}\\
Our study sheds light on the significant barriers older adults encounter when attempting to configure and manage the security and privacy settings of IoT health systems. Despite the benefits our study tells a clear story where our participants have acknowledged that configuring IoT security is more than just a technical task-- for many it's a deeply frustrating experience, making them feel excluded from the very technologies meant to support them. 
While prior research like Yusif et al. ~\cite{YUSIF_2016_adoption}, have shown usability issues faced by older adults in general, our findings specifically highlight the distinct intersection of cognitive overload and IoT security settings.
For many participants, the challenges began with the sheer complexity of IoT systems. $P6$ summarized this struggle well, stating: \lq\lq The new technology is surpassing the ability of seniors to keep up with it.\rq\rq~ This reflects the cognitive overload that older adults often face when navigating unfamiliar interfaces, particularly those filled with complex technical jargon. Our participant, $P3$ also mentioned that \lq\lq it is not just a matter of learning how the settings work--understanding them requires mental energy that many older adults find exhausting.\rq\rq~

Such cognitive burden is further compounded by poorly designed user interfaces. Participants frequently highlighted how confusing and tedious privacy settings were to engage with. $P7$ spoke directly to this frustration, noting: \lq\lq I get fatigued having to read three, four pages of privacy policies all words like, no way.\rq\rq~ This statement captures the essence of a broader issue: the interfaces and security settings of IoT devices are not designed with older adults in mind. When privacy settings become a test of endurance rather than a user-friendly feature, it alienates users and diminishes their sense of control. As these obstacles pile up, the frustration they cause does more than just inconvenience users—it breeds a sense of disempowerment. Older adults know that their inability to configure devices properly increases their vulnerability. As $P1$ pointed out, \lq\lq We know these things can be hacked, but they don't make it easy for us to protect ourselves.\rq\rq~ This feeling of being unable to take proper security measures fosters a dangerous cycle: older adults may avoid using security features altogether, inadvertently leaving themselves more exposed to cyber threats. 

The implications of these findings are significant. Not only do these barriers increase the frustration and anxiety of older adults, but they also elevate the risks of security breaches. Poorly configured devices create an open door for potential attacks, widening the ``attack surface" for older users. The very systems that promise to support independent living instead become sources of fear and discomfort. This sense of vulnerability, driven by technical barriers, is a critical issue that needs to be addressed if IoT is to fulfil its promise for older populations. These findings underscore the urgent need for IoT companies to rethink how they design user interfaces and privacy settings. Interfaces that are overly complex or wordy leave older adults feeling left behind. It's clear that current systems fail to accommodate the specific needs of this demographic. By simplifying privacy controls, providing clearer instructions, and creating more accessible support systems, IoT companies can reduce the cognitive load on older users and help them feel empowered to manage their security. Ultimately, our study highlights that addressing these barriers is not just about improving user experience but also about ensuring the safety and well-being of older adults as they integrate IoT devices into their lives. If we want to see widespread adoption of IoT among older adults, we must focus on designing systems that are not only secure but also easy to use and understand.

\subsection{Impact of Security and Privacy Breaches on Mental Health and Well-being}

\textbf{RQ3:} \textit{What impact do security and privacy breaches have on the well-being of older adults using IoT?}\\
Our study reveals a profound psychological toll that security and privacy breaches in IoT health systems have on older adults. Our data consistently show that beyond the technical challenges, the emotional and mental impact of these incidents can be severe, often eroding trust in technology and causing significant distress. Our participants frequently reported feelings of paranoia, anxiety, and frustration after experiencing or even learning about security breaches. This anxiety is not abstract; it is deeply rooted in real fears of data misuse and personal vulnerability. When asked to rate their level of concern regarding various IoT security issues, respondents consistently rated their worry about unauthorized access to personal health data at a $\mu$ of $4.36$ out of $5$, making it the top concern. Closely following were fears related to data breaches and leaks ($\mu$ = $4.29$) and unauthorized surveillance ($\mu$ = $4.14$). These high scores across the board demonstrate the significant levels of anxiety among older adults about how their data might be exposed or misused, particularly in health-related contexts where privacy is deeply personal.

These survey findings were strongly echoed in the interviews, where participants vividly recounted the emotional fallout from security breaches. $P6$, for instance, described the aftermath of an identity theft incident as \lq\lq extremely invasive,\rq\rq~ explaining that it not only disrupted their financial life but also had lasting emotional consequences. The violation of trust that comes with such incidents left $P6$ feeling anxious and constantly on guard. Similarly, $P5$ expressed their ongoing discomfort, noting that the risk of scams and fraud resulting from data breaches left them uneasy about using IoT devices. These accounts underscore the emotional weight that older adults attach to data security, where the consequences of breaches extend far beyond technical inconvenience. 
These findings contribute to a growing awareness of the psychological impact of security breaches. While Nicholson et al. reveals similar emotional reactions to general online privacy violations ~\cite{Nicholson_2019_Cyber}, our study shows how these responses are amplified in health-related circumstances due to the perceived sensitivity of the information. This observation underscores the pressing necessity for initiatives aimed at both technical protection and psychological reassurance.
What becomes evident from these interviews is that the psychological impact of security incidents is not just about the breach itself--it's about the long-term erosion of trust and confidence in IoT technologies. For older adults, security breaches are more than a digital annoyance; they compromise a sense of safety and control. Participants repeatedly expressed concerns about their ability to manage the aftermath of such incidents. As $P3$ recounted, the emotional burden of dealing with identity theft and reconfiguring financial accounts was overwhelming, leading to a lasting sense of vulnerability. This ongoing anxiety is compounded by fears that older adults may not have the resources or resilience to handle the fallout from security breaches. Several participants mentioned that the emotional strain from such incidents significantly affected their mental health. The combination of worry about the breach itself, coupled with doubts about their capacity to manage future incidents, left many feeling powerless. This emotional toll is especially concerning in a demographic that increasingly relies on IoT devices to support their health and independence.

Our findings highlight the high emotional cost associated with security and privacy breaches in IoT systems. For older adults, the fear of unauthorized access to personal data creates a sense of vulnerability that impacts their overall well-being. This goes beyond simply losing trust in a device—it affects their relationship with technology as a whole. Without confidence in the security of IoT systems, older adults may withdraw from using technologies that could otherwise significantly improve their quality of life. The psychological toll of these breaches, therefore, cannot be ignored. The emotional distress that follows an incident has the potential to undo the positive benefits that IoT devices offer. Older adults' trust in technology has been proven to be fragile in our collected sample, and once broken, it can become difficult to rebuild leading to disengagement like it has shown for other population~\cite{doyle_2023_fragile, feng_2004_empathy}.

\subsection{Coping Strategies and Support Systems}
\textbf{RQ4:} \textit{How do older adults cope with and seek support for managing the security and privacy concerns of health monitoring IoT devices?}

Our study reveals that older adults actively engage in various strategies to navigate the security and privacy concerns associated with health monitoring IoT devices. While earlier discussions have explored the challenges and emotional toll, it's crucial to highlight how older adults are not passive in the face of these obstacles. Instead, they demonstrated resilience and resourcefulness by seeking support, adapting their use of technology, and finding ways to protect their personal data. For many participants, family members—particularly younger relatives played a vital role in helping them manage the complexities of health monitoring devices. $P1$ described how they routinely rely on their grandson to configure their health tracker, saying: \lq\lq I ask my grandson to help me set up these things because he understands them better.\rq\rq~ This reliance on tech-savvy family members helped our participants feel more secure, but it also created a sense of dependency that some participants found frustrating. While family support can bridge the knowledge gap, it leaves many older adults wishing for more autonomy in managing their devices.

In addition to family, several participants mentioned turning to professional services for assistance with health monitoring devices. $P4$ shared how they use a local tech support service designed for older adults, saying: \lq\lq They come over and help set up my health devices, but I still don't fully understand all the settings.\rq\rq~ This proactive approach shows that older adults are eager to stay on top of their health technology, but they often find that professional support does not address their long-term understanding. While these services solve immediate issues, many participants want more comprehensive guidance to manage their devices independently. A common strategy for dealing with security concerns is selective usage. Several participants have chosen to limit their use of health monitoring devices due to fears about where their personal data might end up. $P9$ reflected this mindset, saying: \lq\lq I don't use the health tracking features anymore because I'm worried about what they do with my data.\rq\rq~ This selective use helps mitigate security risks but comes at the cost of missing out on key health monitoring features that could improve their well-being. This trade-off between protecting personal data and fully utilizing the devices reveals an important challenge—older adults are actively making decisions to safeguard themselves, but often at the expense of their health benefits.

Some older adults cope by adopting a resigned outlook toward data privacy. $P7$ expressed this sentiment: \lq\lq I know they have my data, but there's not much I can do about it.\rq\rq~ This perspective, though passive, reflects a realistic approach where older users continue using their health monitoring devices despite their concerns. This resignation suggests that some older adults feel powerless to fully protect their data but have decided to prioritize the health benefits of the devices over the perceived risks. Peer support also played a critical role in how older adults cope with the challenges of managing health monitoring IoT devices. Many participants shared how they discuss these issues with friends or neighbors who face similar concerns. $P3$ explained: \lq\lq We talk about these health gadgets in our community group, and it's comforting to know I'm not the only one who's confused.\rq\rq~ This sense of community creates an informal support network where older adults can share experiences, advice, and reassurance. These peer interactions provide both emotional and practical support, allowing them to manage their health technology more effectively.

These coping strategies show that older adults are not merely enduring the challenges of health monitoring IoT devices—they are actively seeking ways to adapt and protect themselves. Whether through family, professional services, selective usage, or community support, they are finding ways to balance their health needs with privacy concerns. However, these strategies also highlight significant areas for improvement. By simplifying device interfaces, offering clearer guidance, and creating more transparent data management practices, IoT companies can empower older adults to use their health monitoring devices with confidence and independence.

\section{Implications}
Our study provides crucial insights into designing IoT systems tailored for older adults, particularly in the context of health monitoring and independent living. By focusing on security and privacy by design and default, we outline key implications for IoT developers, healthcare providers, and policymakers to enhance the usability, accessibility, and trust in IoT systems for this demographic.

\subsection{Simplifying IoT for Health Monitoring}
The complexity of security settings in health monitoring IoT devices—such as wearables for tracking vitals, medication reminders, or fall detection systems—poses a significant challenge for older adults. As $P6$ noted, \lq The new technology is surpassing the ability of seniors to keep up with it.\rq~ This creates a barrier to adoption and effective use, especially in devices critical for independent living.
Building on Cavoukian's privacy by design framework ~\cite{Cavoukian_2010_PbD}, IoT systems must prioritize \textit{usable security by design}, ensuring that health data remains secure without requiring complex configurations. 
Following Siddiqui et al.'s principle of adaptive security ~\cite{Siddiqui_2023_adaptive}, minimalist interfaces should be implemented, featuring one-click privacy settings and automatic security updates. For instance, health monitoring devices could come with pre-configured best practices, eliminating the need for users to manually adjust intricate settings. Integrating voice-activated commands and biometric authentication, such as facial recognition or fingerprint scanning, can help older adults securely access their devices without the stress of remembering passwords, which can be particularly beneficial for devices that monitor critical health information. Such design changes will not only enhance security but also lower the cognitive burden for older adults, enabling them to manage their health data more effectively. These improvements align with the principles of security and privacy by default, ensuring IoT systems are safe and easy to use from the moment they are activated.

\subsection{Empowering Older Adults with Real-Time Control Over Health Data}
IoT devices for health monitoring collect vast amounts of sensitive information, including heart rates, medication schedules, and activity levels. Participants expressed concerns about who can access this data and for what purpose, with $P9$ stating, \lq I want to know who is seeing my data and why.\rq~ This highlights the need for transparency and control over data-sharing practices, which is often lacking in current IoT ecosystems. To address these concerns, we recommend integrating \textit{real-time data transparency} features that notify users every time their health data is accessed or shared. These notifications should be concise, non-technical, and customizable to meet the preferences of older adults. A privacy dashboard could be included, allowing users to control and review data-sharing settings from a single, easy-to-navigate interface. For example, the dashboard could clearly show who has access to their health data—whether it's a doctor, caregiver, or third-party service—and explain why the data is being shared. This increased level of transparency will empower older adults, allowing them to maintain autonomy over their personal information while ensuring that their health data is only shared when absolutely necessary and with their consent.

\subsection{Seamless Security Updates for Ongoing Protection}
Many older adults in our study mentioned they find it difficult to keep up with security updates, which are vital for protecting IoT devices from emerging threats. As some participants noted, they often miss or avoid these updates due to confusion or lack of awareness. This is especially concerning for health-monitoring devices, where outdated software could leave users vulnerable to cyberattacks.
We advocate for \textit{invisible security updates}—automatic updates that occur seamlessly without requiring user intervention. These updates should include simplified explanations of what has changed, ensuring older adults understand the importance of the update without feeling overwhelmed by technical jargon. Pairing these updates with visual indicators, such as a security badge or rating system, would allow users to easily monitor the security status of their devices at a glance. For example, a heart-monitoring device could display a green badge indicating that it is up-to-date and secure, similar to energy efficiency ratings in household appliances ~\cite{Knayer_2022_EvaluationOR, USEnergy_2024_Ratings}. This system would provide reassurance for older adults, giving them confidence that their health-monitoring devices are continuously protected without adding any extra complexity.

\subsection{Creating a Senior-Friendly IoT Certification for Health and Independent Living}
To ensure older adults can make informed decisions about the security of IoT devices, particularly those used for health monitoring and independent living, we propose the creation of a \lq Senior-Friendly IoT\rq~ certification. This certification would guarantee that a device meets specific security and privacy benchmarks, such as strong encryption, user-friendly interfaces, and transparent data-sharing controls. Much like energy efficiency ratings ~\cite{Knayer_2022_EvaluationOR, USEnergy_2024_Ratings}, this certification would serve as a trusted signal to older adults that the device has been designed with their security, privacy, and usability needs in mind. A certified health-monitoring device, for example, would not only provide clear data-sharing options but also ensure that sensitive health information is protected with industry-standard encryption protocols. This certification could also guide manufacturers to prioritize the needs of older adults during product development, encouraging the industry to shift toward more secure and accessible designs. This initiative could increase the trust and adoption of IoT devices among older adults, particularly those managing chronic conditions or relying on health-monitoring devices to maintain their independence.

\subsection{Strengthening Policy for IoT Health Monitoring Devices}
While design improvements are crucial, our findings indicate a need for policy-level interventions to protect older adults' data in health-monitoring IoT devices. As $P1$ remarked, \lq I don't feel safe unless I know my data is protected by law.\rq~ This underscores the importance of stronger data protection regulations specifically for IoT devices used in healthcare and independent living contexts. Mandatory security and privacy standards should be established further, including encryption protocols, user consent for data sharing, and regular security audits. For example, HIPAA regulations should be expanded to include IoT systems that handle health data, ensuring that these devices meet the same stringent privacy requirements as traditional medical records. Additionally, policymakers should introduce consumer protection measures that require clear, easy-to-understand explanations of how health data is used and shared. Public education campaigns could also help older adults better understand their digital rights and the steps they can take to protect their privacy. Such regulatory frameworks would create a safer IoT environment for older adults, ensuring that health-monitoring devices contribute to independent living without compromising their security or privacy.

\subsection{Co-Creating Secure and Usable IoT Devices for Older Adults}
Our study emphasizes the need for~\textit{collaborative efforts} between IoT developers, healthcare providers, older adult advocacy groups, and policymakers to create secure and user-friendly IoT systems for health monitoring and independent living. Co-design sessions with older adults should be an integral part of the development process, ensuring that their specific needs are addressed.
Collaborative efforts can ensure that IoT devices are both technologically robust and tailored to the cognitive and physical capabilities of older adults. For example, healthcare providers could work with developers to ensure that health-monitoring devices are designed with simple, intuitive interfaces that cater to older adults' comfort levels with technology. Ongoing forums between these stakeholders would allow for continuous feedback and iterative improvements, keeping pace with the evolving needs of older adults and the rapid advancement of IoT technologies. This approach would not only improve the security and usability of IoT devices but also foster greater trust and adoption, ultimately contributing to better health outcomes and enhanced quality of life for older adults.

\section{Limitations and Future Work}
We provide valuable insights into older adults' perceptions of IoT security, privacy, and health monitoring. Although our sample size of $22$ survey participants and $9$ interviewees may appear small, it was sufficient to achieve data saturation for our exploratory analysis. The study's findings offer a strong foundation for understanding the key issues, despite recruitment challenges common when working with older populations. 
Building on the insights of Knowles and Hanson~\cite{Knowles_2018_Tech}, socioeconomic factors are known to play a critical role in shaping both access to and attitudes toward technology. However, this study did not account for socioeconomic status which may have influenced participants' perceptions of privacy and security, potentially shaping their experiences and responses. 
Our focus on participants from the United States helped maintain control over variability, though it may limit generalizability to other cultural or regulatory environments. In future work, our aim will be to recruit larger and more diverse samples across geographic and socioeconomic backgrounds to enhance generalizability. Additionally, we will focus on designing and evaluating user-centered security features tailored to the specific cognitive and physical needs of older adults, ensuring these solutions are adaptable across diverse contexts.

\section{Conclusion}

Our study highlights the emotional and practical challenges older adults face when interacting with IoT, particularly in health monitoring. While IoT technologies offer valuable support for independent living, the security and privacy concerns reported by our participants highlight significant barriers to trust and full engagement. Many of our participants expressed a mix of anxiety and frustration over the complexity of security settings, with one participant sharing their confusion over data-sharing policies and concerns about unauthorized access to sensitive health data. Despite $91\%$ of our participants saying they have used IoT devices for health monitoring, only $14\%$ of them felt the security and privacy measures were \lq\lq very effective\rq\rq~, and more than half ($63.64\%$) rarely even review or carry out updates on their IoT devices.
 After experiencing a data breach, one participant described a heightened sense of vulnerability and ongoing mistrust in the technology, reflecting the psychological toll these security issues impose. In addition, participants found it difficult to navigate security settings, and some relied on younger family members or external help to cope with the complexity, which often led to disengagement with devices altogether.
 Participants overwhelmingly recommended real-time notifications of data access and simpler security interfaces. These insights call for IoT systems to incorporate adaptive security, real-time data transparency, and automated updates to better support older adults, particularly those using IoT devices for health monitoring and independent living. By addressing these needs, IoT technologies can enhance trust and guarantee that older adults can derive benefits from these health monitoring devices without compromising their privacy or experiencing technical complexity.

\bibliographystyle{ACM-Reference-Format}
\bibliography{ref}

\clearpage
\appendix
\section*{APPENDIX}
\section{Survey Questionnaires}
\label{app:survey}

This appendix presents the complete questions used in our survey on IoT security and privacy concerns among older adults. The questions include pre-screening, device usage, awareness of security features, privacy concerns, and demographic information.

Throughout the survey, participants were asked to respond to questions using various response formats:
\begin{itemize}
    \item For questions assessing \lq\lq level of agreement\rq\rq, participants used a 5-point Likert scale ranging from "Strongly Disagree" to "Strongly Agree".
    \item For questions about \lq\lq frequency\rq\rq, participants chose from options such as "Daily", "Weekly", "Monthly", "Rarely", or "Never".
    \item For questions assessing \lq\lq concern levels\rq\rq, participants used a 5-point scale ranging from "Not Concerned" to "Very Concerned".
    \item Some questions were \lq\lq  multiple-choice\rq\rq~ with predefined options, while others allowed for \lq\lq open-ended responses\rq\rq~ where participants could elaborate on their experiences.
\end{itemize}

\begin{enumerate}
    \item What is your age?
    \item Are you currently residing in the US?
    \item Do you commit to answering the questions in this survey honestly?
    \item Have you used IoT or similar wearable devices for health monitoring and activity tracking in the past 12 months?
    \item If you have used IoT, how concerned are you about the security and privacy of your personal information when using the device?
    \item Please rate the following security and privacy concerns regarding IoTs on a scale of 1 to 5, where 1 represents "Not Concerned" and 5 represents "Very Concerned":
    \begin{itemize}
        \item Unauthorized access to personal health data
        \item Data breaches and leaks
        \item Tracking and monitoring by unauthorized individuals
        \item Inadequate data encryption
        \item Lack of transparency in data handling practices
    \end{itemize}
    \item Are you aware of the security and privacy features offered by IoT, such as two-factor authentication, data encryption, and privacy settings?
    \item Based on your experience or knowledge, how effective do you think IoT's security and privacy measures are in protecting your personal information?
    \item Please rate your level of agreement regarding IoT security and privacy on a scale of 1 to 5, where 1 represents "Strongly Disagree" and 5 represents "Strongly Agree":
    \begin{itemize}
        \item IoT adequately informs me about the security and privacy features of their devices.
        \item I trust IoT to handle my personal health data securely.
        \item I feel confident in the data encryption measures employed by IoT.
        \item IoT provides sufficient control over the sharing and access of my personal health data.
    \end{itemize}
    \item How frequently do you review and update the security and privacy settings on your IoT device?
    \item In your opinion, what are the most significant challenges older adults face when trying to ensure the security and privacy of their IoT data?
    \item Have you ever encountered any security or privacy-related incidents while using IoT or similar device? If yes, please briefly describe the incident and its impact on you.
    \item How aware are you of the potential security risks associated with using third-party applications or services that integrate with IoT?
    \item How confident are you in identifying and reporting potential security or privacy threats related to your IoT device?
    \item Have you ever taken any specific actions to enhance the security and privacy of your IoT device or personal health data? If yes, please describe those actions.
    \item How comfortable are you with sharing your personal health data collected by IoT with healthcare providers or researchers for the purpose of improving healthcare services or research?
    \item Do you have any concerns regarding data collection practices in IoT devices?
    \item Have you ever disabled or stopped using an IoT device due to security or privacy concerns? If yes, why?    
    \item How important is it for you to have control over the data that IoT devices collect about you?
    \item Would you be willing to pay extra for IoT devices with stronger security and privacy features?    
    \item Do you believe that IoT device manufacturers provide enough transparency regarding how they handle and store your data?
    \item Have you received any guidance or training on how to manage the security and privacy settings on your IoT devices?   
    
    \item How do you typically utilize your IoT device in your daily life?
    \item How likely are you to recommend IoT devices to other older adults based on their security and privacy features?
    \item What sources of information do you rely on to stay informed about the security and privacy of your IoT device and related services?
    \item Overall, how satisfied are you with the security and privacy measures provided by IoT for their devices?
    \item What additional security and privacy features or improvements would you like to see in IoTs to address your concerns and protect your personal information?
    \item What is the highest level of education you have completed? (if currently enrolled, highest degree received.)
    \item What gender do you identify as?
    \item Please specify your ethnicity.
    \item If you have anything to add or share with us related to the topic, please feel free to write it here.
\end{enumerate}

\section{Semi-Structured Interview Questions}
\label{app:interview}

This appendix presents the set of interview questions used in our study to explore IoT security and privacy concerns among older adults. 

\begin{enumerate}

    \item Can you describe your experience using IoT or smart devices for health monitoring and independent living?
    \item What motivated you to start using IoT or smart devices for health monitoring and independent living?
    \item What security and privacy concerns, if any, do you have when using your IoT or smart devices?
    \item How important are these security and privacy concerns to you when using IoT or smart devices?
    \item Have you encountered any specific situations or incidents that raised security or privacy concerns while using your IoT or smart devices? If yes, could you describe them?
    \item How do you perceive the benefits of using IoT or smart devices for health monitoring and independent living compared to the potential risks associated with security and privacy?
    \item Are you aware of the security and privacy features provided by IoT or smart devices?
    \item How comfortable do you feel using the security and privacy features provided by IoT or smart devices to protect your personal information?
    \item In your opinion, what are the most critical security and privacy challenges faced by older adults when using IoT or smart devices?
    \item How do these security and privacy challenges impact your overall experience with IoT or smart devices?
    \item How well do you think IoT or smart devices address the security and privacy concerns specific to older adults?
    \item What improvements or additional security and privacy features would you like to see in IoT or smart devices?
    \item What are your expectations regarding the protection of your personal health data collected by IoT or smart devices?
    \item How confident are you in IoT's data handling practices, such as data anonymization methods and storage?
    \item Do you feel adequately informed about the data collection practices of IoT or smart devices?
    \item How transparent do you perceive IoT or smart devices to be regarding data collection, storage, and sharing practices?
    \item What recommendations would you provide to enhance the security and privacy of IoT or smart devices for older adults?
    \item Are there any specific features or changes that you believe would better align IoT or smart devices with your security and privacy needs?

\end{enumerate}

\end{document}